\documentclass[floatfix,groupedaddress,superscriptaddress,aps,amsmath,amssymb, pra,longbibliography, twocolumn,a4]{revtex4-1}
\usepackage[amssymb]{SIunits}
\usepackage{mathrsfs}
\usepackage{graphicx,float}
 \usepackage{subfigure}
\usepackage{color}

\begin{document}
\title{Quantum routing of single optical photons with a superconducting flux qubit}

 \author{Keyu Xia}  %
 \email{keyu.xia@mq.edu.au}
 \affiliation{Centre for Engineered Quantum Systems, Department of Physics and Astronomy, Macquarie University, NSW 2109, Australia}

 \author{Fedor Jelezko}
 \affiliation{Institute for Quantum Optics and Center for Integrated Quantum Science and Technology (IQst), Ulm University, Ulm D-89081, Germany}

 \author{Jason Twamley}
 \affiliation{Centre for Engineered Quantum Systems, Department of Physics and Astronomy, Macquarie University, NSW 2109, Australia}

\date{\today}

\begin{abstract}
  Controlling and swapping quantum information in a quantum coherent way between the microwave and optical regimes is essential for building long-range superconducting quantum networks but extremely challenging. We propose a hybrid quantum interface between the microwave and optical domains where the propagation of a single-photon pulse along a nanowaveguide is controlled in a coherent way by tuning electromagnetically induced transparency window with the quantum state of a flux qubit. The qubit can route a single-photon pulse with a single spin in nanodiamond into a quantum superposition of paths without the aid of an optical cavity - simplifying the setup. By preparing the flux qubit in a superposition state our cavity-less scheme creates a hybrid state-path entanglement between a flying single optical photon and a static superconducting qubit, and can conduct heralded quantum state transfer via measurement.
\end{abstract}

\maketitle

\section{Introduction}
Quantum networks are an essential component for scalable quantum information processing and quantum communication \cite{QNetworkKimble,QNetworkDuan}. A key element to build a quantum network is a quantum router \cite{QRouterNet1,QRouterNet2,QRouterNet3}, which coherently communicates between distant quantum nodes using photons. A quantum router determines the outgoing channel of the input flying  photons by the quantum state of a static control qubit and this must be achieved in a coherent fashion.

Solid state qubits like superconducting qubits (SQs) working in the microwave (mw) domain are perhaps the most promising candidate for scalable quantum computation. However,  communicating between remote SQs requires the transport of optical photons. A quantum interface bridging the mw and optical domains has been proposed based on optomechanical transduction \cite{OptomMWOInterfaceAClerk,OptomMWOInterfaceLTian,OptomMWOInterfaceMilburn,OptomMWOInterfaceDuan,OptomMWOInterfaceLukin,OptomMWOInterfaceJason}, frequency mixing in ensembles of spins \cite{SpinMWOInterfaceFleischhauer,SpinMWOInterfaceLongdell, SpinMWOInterfaceJason} or atoms \cite{AtomicMWOInterfaceTaylor,AtomicMWOInterfaceKiffner}. So far, all these works require the transfer of excitations  between the mw and optical domains and usually requires large magnetic coupling \cite{OptomMWOInterfaceAClerk,OptomMWOInterfaceLTian,OptomMWOInterfaceMilburn,OptomMWOInterfaceDuan,OptomMWOInterfaceLukin,OptomMWOInterfaceJason,SpinMWOInterfaceFleischhauer,SpinMWOInterfaceLongdell, SpinMWOInterfaceJason,AtomicMWOInterfaceTaylor,AtomicMWOInterfaceKiffner}. In contrast, the quantum router can be more advantageous for quantum networks \cite{QRouterNet1,QRouterNet2,QRouterNet3}, since it  creates  state-path entanglement between a flying photon and a static qubit. So far, quantum routers can only work in the mw domain \cite{mwRouterChongLi,mwRouterCPSun,mwRouterHoi} or the optical domain \cite{ORouterMFeng,ORouterHengFan,ORouterJason1,ORouterJason2,ORouterKimble,ORouterDayan,ORouterDuan,ORouterArno}, separately. However, a key challenge for SQ-based quantum networks  is to achieve the hybrid quantum routing of optical photons by qubits working in the mw domain.

Here we present a scheme to route a single-photon pulse into a quantum superposition of output paths by a quantum magnetic field generated by a flux qubit. Our cavity-less scheme also create hybrid entanglement between the propagation paths of a flying optical photon and the states of a static superconducting qubit. Such entanglement has only been demonstrated recently by using cavity QED in the optical domain \cite{QCPFGate1}. Our scheme is able to route the optical photons with little change in pulse shape and does not directly exchange excitation between the static and flying qubits. Our scheme is tailored for a superconducting qubit - a flux qubit in particular, and does not require the demanding integration of high-$Q$ optical cavities. In addition, in contrast to previous works making use of  off-resonant strong coupling to swap excitation, the driving in our scheme is nearly on-resonance and more efficient. our method only requires a weak magnetic spin-flux qubit coupling larger than the decoherence of the ground states of the single spin and the flux qubit.
More importantly, we route the single photon by tuning an Electromagnetically Induced Transparency (EIT) window of a single spin. This method only requires a magnetic coupling larger than the decoherence of the ground states of the single spin and the flux qubit.

\section{Model}
\subsection{System}
\begin{figure}[h]
 \centering
 \includegraphics[width=0.9\linewidth]{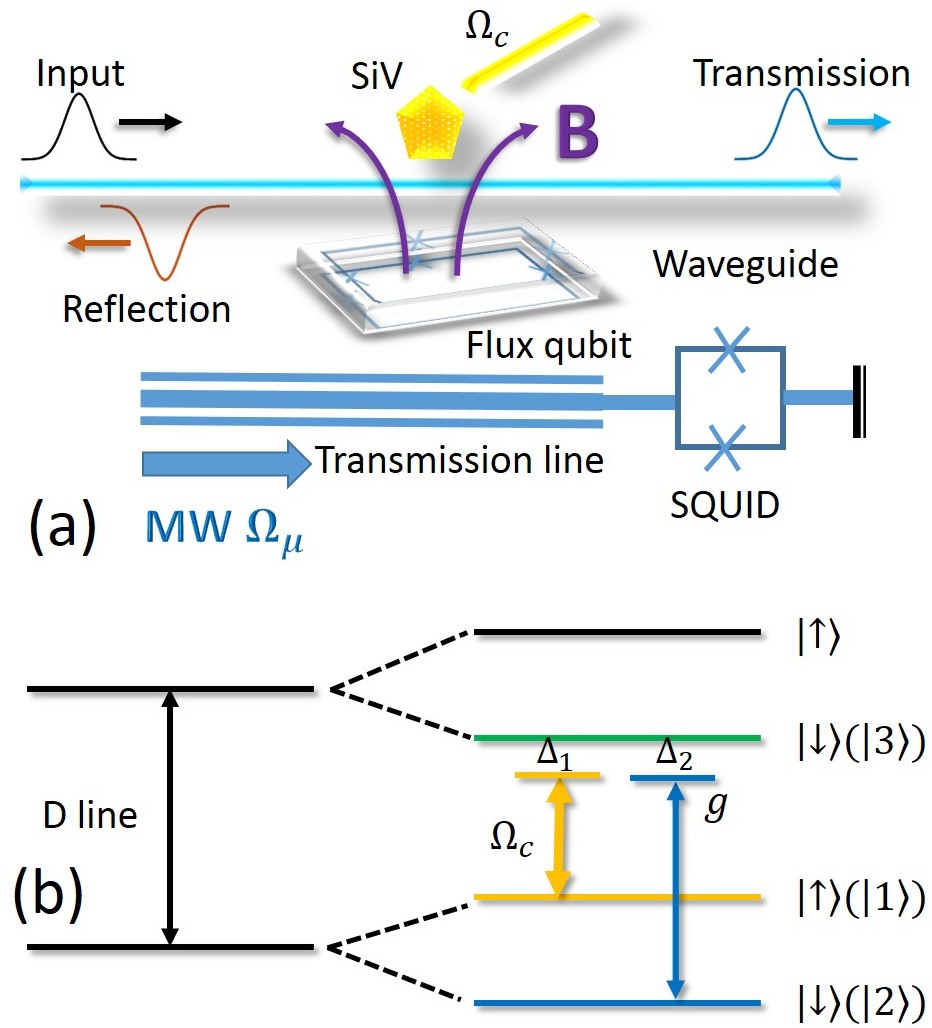} \\
 \caption{(Color online) (a) Schematic of a hybrid quantum single-photon router at $20~\text{\milli\kelvin}$. A single-photon pulse with carrier frequency of $\omega_\text{in}$ (black pulse) inputs to a nanowaveguide, e.g. a nanofiber or a photonic crystal waveguide, from the left. It couples to a nearby single SiV with strength $g$. An external coherent laser field $\Omega_c$ drives a transition in the SiV and creates an EIT window. The input pulse partly passes through the SiV (blue pulse) and is partly scattered backwards (red pulse). The central frequency of the EIT window is modulated by the energy levels of the SiV, which themselves are shifted due to the quantum magnetic field $B$ (purple lines) dependent on the quantum state of the flux qubit, $|e\rangle$ and $|g\rangle$. A classical microwave field, $\Omega_\mu$, from a nearby SQUID-terminated transmission line  \cite{MWTransLine1,MWTransLine2}, modulates the flux qubit energies quickly in time (b) Level diagram of the SiV. The SiV acts as a $\Lambda$-type configuration with the excited state $|3\rangle$, and two ground states $|2\rangle$ and $|1\rangle$ \cite{FedorSiVEIT,BenjaminSiVEIT}.} \label{fig:system}
\end{figure}


The main idea for our quantum routing of a single traveling photon with a flux qubit is depicted in Fig. \ref{fig:system}. A single photon pulse propagating in a nanowaveguide is dominantly scattered forward and backward by a $\Lambda$-type three-level ``atom'' like silicon vacancy defect (SiVs) in nanodiamond. Our scheme requires the strong optical coupling regime and such large coupling strengths in a waveguide setup can be achieved using various nanosctructures \cite{LargePurcellNanofiber,LargePurcellSlotwg,LargePurcellPCwg1, LargePurcellPCwg2, LargePurcellPCwg3, LargePurcellPCwg4, LargePurcellPCwg5,LargePurcellPlasmon}. In our proposal the atom acts as a beam splitter to control the transmission and reflection of the input single-photon pulse. An external classical laser pulse, $\Omega_c$, is applied to open an electromagnetically induced transparency (EIT) window for the input single photon. For a long input single-photon pulse nearly resonant with the transition of $|2\rangle \leftrightarrow |3\rangle$, under the condition of two-photon resonance, the atom is transparency and the single photon can remain right-moving, whereas the single photon is completely reflected backward when two-photon detuning is large but the single photon is nearly resonant with the atom. Therefore, shifting the energy levels of the atom can control the transmission and reflection of the incident single photon. To do so, a magnetic field $\mathbf{B}$ created by a flux qubit is applied to shift the levels of $|1\rangle$, $|2\rangle$ and $|3\rangle$. This quantum magnetic field is dependent on the quantum state of the flux qubit. As a result, one can use the flux qubit to route the input single photon to a superposition state of two output paths. Essentially, this device can create  hybrid entanglement between a flying optical photon and a static superconducting qubit.

We assume that the involved three levels, $|j\rangle$, of SiV have energies $\omega_j$ with $j\in \{1,2,3\}$. The coherent laser field with frequency $\omega_\text{c}$ and the traveling single photon with carrier frequency $\omega_\text{in}$, which drives the transitions $|1\rangle \leftrightarrow |3\rangle$ and 
$|2\rangle \leftrightarrow |3\rangle$ with detunings $\Delta_1 = (\omega_3 - \omega_1 -\omega_\text{c})$ and $\Delta_2 = (\omega_3 - \omega_2 -\omega_\text{in})$, respectively. These levels move due to the quantum state of the flux qubit through linear energy shifts $\pm \eta_j \sigma_\text{z}^\prime=\mu_B g_{\text{e},j} \mathbf{B} \sigma_\text{z}^\prime$ with $j\in \{1,2,3\}$, where $\mu_B = 14~\text{\giga\hertz}/\text{T}$,  $\mu_B$ is the Bohr magneton, $g_{\text{e},j}\approx 1$ is the g-factor of the corresponding level for spin-$\frac{1}{2}$, $|e\rangle$ and $|g\rangle$ are the excited and ground states of the flux qubit, and $\sigma_\text{z}^\prime = |e\rangle \langle e| - |g\rangle \langle g|$. We arrange that the diamond is cut along the $z$ direction and $\mathbf{B}$ orients to $z$ as well. In such arrangement, the energy shifts of SiV due to the flux qubit is given by $(\eta_1 S_{11}-\eta_3 S_{33}-\eta_2 S_{22}) \sigma_{z}^\prime$ with $S_{lj} = |l\rangle \langle j|$ and $l, j \in \{1, 2, 3\}$.   When the flux qubit is prepared in state $|e\rangle$, the $j$\emph{th} level is shifted up ($j=1$) or down ($j=2,3$) by $\eta_j$, whereas the shift is reversed to $-\eta_j$ for $|g\rangle$. The real two-photon detuning is $\delta = \Delta_1 - \Delta_2 - (2\eta_3+\eta_1-\eta_2)\,\sigma_z^\prime$. The merit we use a SiV center is that it is expected to have very long decoherence time, $T_2$, for spin at \milli\kelvin temperatures \cite{NJP17.043011}.

A classical magnetic field $\Omega_\mu$ is applied to prepare the initial state, and then rapidly modulates the energy levels  of the flux qubit through a transmission line (TL), which is terminated by a Superconducting Quantum Interference Device (SQUID). The flux qubit can also decay via the coupling to this transmission line treating the latter as an environment. The flux qubit decays at a rate of $\gamma_f$ and has a pure dephasing rate $\Gamma^*$. We propose to control the flux qubit via the SQUID-terminated TL as one can  dynamically tune the coupling of the flux qubit to the transmission line by varying the terminating boundary condition by tuning the SQUID. Once the flux qubit is prepared to the desired state, we can decouple it from the TL environment thus decreasing the flux qubit's decay rate to vanishing, by tuning the flux threading  through the loop of SQUID \cite{DecouplefromEnvorn1,DecouplefromEnvorn2,DecouplefromEnvorn3}. 

We first derive the Hamiltonian governing the dynamics of the presented hybrid quantum system. The flux qubit, with excited state $|e\rangle$ and ground state $|g\rangle$, creates a quantum magnetic field $\mathbf{B}$ dependent on its inner state. The transition frequency of this flux qubit can be tuned with a bias flux. If the bias flux includes a weak continuous microwave field $\Omega_\mu$ oscillating at frequency $\omega_\mu$, the Hamiltonian for the flux qubit is 
\begin{equation} \label{eq:Hfluxorigin}
 H_\text{flux} = \frac{\in}{2}\sigma_z^\prime + \frac{\mathscr{T}}{2}\sigma_x^\prime + \Omega_\mu \cos(\omega_\mu t) \sigma_z^\prime\;,
\end{equation}
where $\sigma_x^\prime=|e\rangle \langle g| + |g\rangle \langle e|$ is the spin operator for spin-$\frac{1}{2}$, $\in$ is the energy difference between $|e\rangle$ and $|g\rangle$, and $\mathscr{T}$ is the tunnel between these two states. For a flux qubit used here, $\in=2I_p(\Phi_b - 0.5\Phi_0)$ is determined by the persistent current $I_p$ circulating along the loop of the flux qubit and the flux bias $\Phi_b$ \cite{TOrlandoFluxQubit}. $\Phi_0$ is the quantum flux. At the so-called sweet point, $\in=0$ and the flux qubit has the longest coherence time. $\mathscr{T}$ is normally a few GHz. Then in the dressed basis of $|\Psi_\pm\rangle = (|e\rangle \pm |g\rangle)/\sqrt{2}$, we can rotate the frame as $\sigma_z^\prime \rightarrow\sigma_x$ and $\sigma_x^\prime \rightarrow\sigma_z$ \cite{TOrlandoFluxQubit}. The Hamiltonian Eq. (\ref{eq:Hfluxorigin}) in this rotated coordinate system becomes
\begin{equation}
 H_\text{flux} = \frac{\mathscr{T}}{2}\sigma_z + \Omega_\mu \cos(\omega_\mu t) \sigma_x \;.
\end{equation}
Now the microwave field becomes a driving. On the resonance, $\omega_\mu=\mathscr{T}$, we can rotate the frame back to the bare basis of $\{|e\rangle, |g\rangle \}$. In this basis, $\sigma_z \rightarrow\sigma_x^\prime$ and $\sigma_x \rightarrow\sigma_z^\prime$ \cite{TOrlandoFluxQubit}, and the Hamiltonian becomes
\begin{equation}
 H_\text{flux} = \frac{\Omega_\mu}{2} \sigma_z^\prime \;.
\end{equation}
When the mw field is very weak that $\Omega_\mu$ is much smaller than the bandwidth of the input single-photon pulse, then we can neglect the coherent motion of the flux qubit (setting $\Omega_\mu \rightarrow 0$) but only consider its decoherence.

The waveguide mode can be either left-moving or right moving. The notations $C_\text{R}^\dag(x)$ and $C_\text{L}^\dag(x)$ indicate the creation of a right- or left-moving photon at position $x$. As shown in Fig. \ref{fig:system}, the $\Lambda$-type three-level system like SiV at $x=0$ interacts with both the left- and right-moving photons with carrier frequency $\omega_\text{in}$. These waveguide modes drives the transition of $|3\rangle \leftrightarrow |2\rangle$ with a coupling rate of $g$. At the same time, the transition of $|3\rangle \leftrightarrow |1\rangle$ is driven by an extra coherent laser field  $\Omega_c$ with frequency $\omega_c$. The excited state $|3\rangle$ of SiV decays to the ground state $|1\rangle (|2\rangle)$ with a rate of $\gamma_1 (\gamma_2)$. While the decay, $\Gamma$, from $|1\rangle$ to $|2\rangle$ is negligible small. Based on the above description, 
%
%
the Hamiltonian describing the interaction between the propagating photon and the ``static'' subsystem is
\begin{equation} \label{eq:Hmoving}
\begin{split}
 H^{(1)} = & [(\Delta_2\pm\eta_2 \mp\eta_3)-i(\gamma_1 + \gamma_2)]S_{33} \\
 & + (\Delta_2-\Delta_1 \pm \eta_1 \pm \eta_2)S_{11} + \Omega_c (S_{31} + S_{13}) \\
 & -iv_g \int dx\, C_\text{R}^\dag(x) \frac{\partial}{\partial x} C_\text{R} + iv_g \int dx\, C_\text{L}^\dag(x) \frac{\partial}{\partial x} C_\text{L} \\
 & + g \int dx \,\delta(x) [(C_\text{R}^\dag+C_\text{L}^\dag)S_{23} + (C_\text{R}+C_\text{L})S_{32}] \;,
\end{split}
\end{equation}
where $v_g$ is the velocity of the light in the nanowaveguide. The upper(lower) signs in front of $\eta_j$ in (\ref{eq:Hmoving}), correspond to the flux states $|e\rangle (|g\rangle)$. The decay of state $|1\rangle$ of SiV is negligible at $20~\text{\milli\kelvin}$. The right- and left-moving wave packets of photons can be written as $|\Phi_R\rangle = \tilde{\phi}_R(x)C_\text{R}^\dag(x) |\varnothing\rangle$ and  $|\Phi_L\rangle = \tilde{\phi}_L(x)C_\text{L}^\dag(x) |\varnothing\rangle$ with $|\varnothing\rangle$ is the vacuum state of the waveguide mode. The corresponding excitations in the right- and left-moving modes are given by $e_R=\int dx |\tilde{\phi}_R(x)|^2$ and $e_L=\int dx |\tilde{\phi}_L(x)|^2$. We define a contrast measure for routing as $\mathfrak{C}= |e_R-e_L|/(e_R + e_L)$.

\subsection{Steady-state solution}
Now we go to find the steady-state transmission and reflection of the input single photon by using the method developed by Shanhui Fan et al. \cite{ShanMethod1,ShanMethod2}. In Fan's method, the general state can be expressed as
\begin{equation} \nonumber \label{eq:state}
\begin{split}
 |\Phi(t)\rangle = & \left[\int dx \tilde{\phi}_R(x)C_\text{R}^\dag(x) +  \tilde{\phi}_L(x)C_\text{L}^\dag(x) \right] |\varnothing,2\rangle \\
 & \otimes (\alpha |g\rangle + \beta|e\rangle) \\
 & + \left[\tilde{e}_3 |\varnothing,3\rangle + \tilde{e}_1 |\varnothing,1\rangle \right] \otimes (\alpha |g\rangle + \beta|e\rangle) \;,
 \end{split}
\end{equation}
where $\tilde{e}_1$ and $\tilde{e}_3$ are the excitation amplitude of the SiV in state $|1\rangle$ and $|3\rangle$, respectively. Tracing over the SiV, we can write the density matrix in a basis with four states, $\{C_\text{R}^\dag(x) |\varnothing,g\rangle, C_\text{R}^\dag(x) |\varnothing,e\rangle $, $ C_\text{L}^\dag(x) |\varnothing,g\rangle, C_\text{L}^\dag(x) |\varnothing,e\rangle\}$, only involving the waveguide mode and the flux qubit.
Using the Schr\"odinger's equation, we can find the amplitudes of transmission and reflection for a single-photon input being 
\begin{subequations} \label{eq:ssTR}
 \begin{align}
  t = & (t_\text{e} +1)/2 \;,\\ 
  r = & (t_\text{e} -1)/2 \;,
 \end{align}
\end{subequations}
with $t_\text{e} =  \frac{\left[  (\Delta_2-\Delta_1) \pm \eta_1 \pm \eta_2\right] \left[\Delta_2 \pm\eta_2 \mp \eta_3+ i\Gamma_\text{wg} - i(\gamma_1+\gamma_2) \right] - \Omega_c^2} {\left[(\Delta_2-\Delta_1) \pm \eta_1 \pm \eta_2\right] \left[\Delta_2 \pm \eta_2 \mp \eta_3- i\Gamma_\text{wg} - i(\gamma_1+\gamma_2) \right] - \Omega_c^2}$. $\Gamma_\text{wg}= g^2/v_g$ is the rate of the SiV to emit photons into the nanowaveguide \cite{ShanMethod1,ShanMethod2}. For simplicity, we assume that $\eta_1=\eta_2=\eta_3=\eta$ for all investigation below.
The transmission and reflection are $T=|t|^2$ and $R=|r|^2$, respectively and these quantities represent the probabilities for the excitation of right- and left-moving scattered wave packets for a right-moving single-photon input. 

\subsection{Cascaded Open System Description}
Fan's scheme is very powerful for studying the dynamics of a system interacting with traveling photons but is hard to take into account the decoherence of the system.  Instead we consider a cascaded master equation with a source cavity injecting a photon into the nanowaveguide-SiV quantum system while also including dephasing of the system \cite{CascadedSystem1,CascadedSystem2}. This method eliminate the explicit waveguide mode in the Hamiltonian and is widely used to study quantum systems with a nonclassical inputs \cite{OptomMWOInterfaceLukin,OptomMWOInterfaceJason,SpinMWOInterfaceJason}. The cascaded master equation method also can be used to evaluate the entanglement and fidelity of the scattered state to be close to particular target state. We use a ``source'' cavity to provide a single-photon pulse. This source is equivalent to the single-photon wave packet in Fan's scheme after applying the relation $x=v_g t$. By assuming a time-dependent decay rate of the source cavity we can generate an arbitrary wave function for the single photon wavepacket incoming to the nanowaveguide-SiV system. The single photon from the source cavity directly drive the transition of $|2\rangle \leftrightarrow |3\rangle$ after a delay, which can be assumed to be zero. Following \cite{CascadedSystem1,CascadedSystem2},  the dynamics of our system can be described by the Hamiltonian
\begin{equation} \label{eq:HO}
  H^{(2)}/\hbar = H_\text{R}+ H_\text{SiV} + H_\text{flux} + H_\text{I} \;,
\end{equation}
with $H_\text{R} = \omega_\text{in}a^\dag a$, $H_\text{flux} = \frac{\Omega_\mu}{2} \sigma_z^\prime$, $H_\text{SiV} = \sum_j \omega_j S_{jj} + \Omega_c \left(e^{-i\omega_c t} S_{31} + e^{i\omega_c t} S_{13} \right)$, and $H_\text{I} =  (\eta_1 S_{11}- \eta_1 S_{22} -\eta_1 S_{33})\sigma_z^\prime$,
and a superoperator linking the cavity and the atom as
\begin{equation} \label{eq:LNet}
 \mathscr{L}_\text{Net}\rho = - \sqrt{\xi_c \kappa_c \xi_2\Gamma_2} (S_{32}a\rho -a \rho S_{32} + \rho a^\dag S_{23}-S_{23}\rho a^\dag) \;,
\end{equation}
with $\xi_c = \frac{\kappa_\text{ex}}{\kappa_c} \leqslant 1$ and $\xi_2=\frac{\Gamma_\text{wg}}{\Gamma_2} \leqslant 1/2$. $\kappa_c$ is the total decay rate of the source cavity, and $\Gamma_2=\gamma_2+ 2\Gamma_\text{wg}$ is the total decay from $|3\rangle$ to $|2\rangle$. $\kappa_\text{ex}$ and $\Gamma_\text{wg}$ are the decay into the waveguide from the source cavity and the SiV, respectively. Note that the SiV decays into two channels: the right- and left-moving modes with rate $\Gamma_\text{wg}$ for each. $\xi_c=1$ and $\xi_2=0.5$ indicate all excitation decays into the nanowaveguide.
$H_\text{R}$ is the cavity model used to generate an arbitrary single-photon pulse in cascade. To create a single photon input pulse in the nanowaveguide we can set this ``source'' cavity initially in the Fock state $|1\rangle$. The wave function of the input single photon can be controlled by a time-dependent decay from this source cavity $\kappa_c(t)$, and we assume $\kappa_c(t)=2\Gamma_\text{wg} e^{-(t-\tau)/2\tau_p^2}$ with a duration of $\tau_p$ and a delay $\tau =5.5\tau_p$ to provide a Gaussian-like single-photon pulse. Such delay is large enough to ensure the waveguide initially in the vacuum state.   $H_\text{SiV}$ describes the free Hamiltonian and the classical driving of the SiV. $H_\text{flux}$ is the Hamiltonian describing the evolution of the flux qubit modulated by the classical mw field. While the energy shifts of SiV due to the flux qubit is given by $H_\text{I}$. 

Under the unitary transformation $U_2= \exp\left\{-i\left[\omega_\text{in} a^\dag a + \frac{\Omega_\mu}{2} \sigma_z^\prime \right] t\right\}$ $\otimes \exp\left\{-i\left[\omega_3 S_{33} + (\omega_3-\omega_\text{in})S_{22} +(\omega_3-\omega_c)S_{11} \right]t\right\}$, the Hamiltonian in Eq. (\ref{eq:HO}) can be rewritten as
\begin{equation} \label{eq:H}
 H^{(2)}= -\Delta_2 S_{22} - \Delta_1 S_{11} + \Omega_c (S_{31} + S_{13}) + H_\text{I} \;.
\end{equation}
The dynamics of the system can be completely described by the master equation
\begin{equation} \label{eq:MEq}
\begin{split}
 \dot{\rho} &=  -i[H^{(2)},\rho] + \mathscr{L}_\text{Net}\rho + \mathscr{L}(\kappa_c,a)\rho + \mathscr{L}(\gamma_1,\sigma_{13})\rho \\
 & + \mathscr{L}(\Gamma_2,\sigma_{23})\rho   + \mathscr{L}(\gamma_f,\sigma_{ge})\rho + \mathscr{L}(\Gamma^*,\sigma_{ee}-\sigma_{gg})\rho\;,
\end{split} 
\end{equation}
where $\mathscr{L}(\gamma,A)\rho=\gamma/2 \{2A\rho A^\dag - A^\dag A \rho- \rho A^\dag A\}$ with $\sigma_{ge} = |g\rangle \langle e|$ and $\sigma_{eg}=\sigma_{ge}^\dag$. $\gamma_f$ and $\Gamma^*$ are the decay and pure dephasing rate of the flux qubit, respectively.
The SiV interacts with both the right- and left-moving photons. Therefore, the transmitted (right-moving) and reflected (left-moving) photons can be determined according to the input-output relation \cite{CascadedSystem1,CascadedSystem2} as 
\begin{subequations} \label{eq:output}
\begin{align}
C_R(x) = & \sqrt{\xi_c\kappa_c(t)}\, a(t) + \sqrt{\xi_2 \Gamma_2}\,\sigma_{32}(t) \;,\\
C_L(x) = & \sqrt{\xi_2 \Gamma_2}\,\sigma_{32}(t) \;.
\end{align}
\end{subequations}
Here we use the fact that $x=v_g t$ and set $v_g=1$. 

The main goal of our scheme is to create an entangled state between the flying photon and the static superconducting qubit. In the basis of $\{C_\text{R}^\dag(x) |\varnothing,g\rangle, C_\text{R}^\dag(x) |\varnothing,e\rangle, C_\text{L}^\dag(x) |\varnothing,g\rangle, C_\text{L}^\dag(x) |\varnothing,e\rangle\}$, we consider the initial state of the input photon in a right-moving single photon, $C^\dag_R(x) | \varnothing, g\rangle$, and the flux qubit in the state of $|\Psi_\text{f,in}\rangle=(\alpha |g\rangle + \beta  |e\rangle)$ with $|\alpha|^2+|\beta|^2=1$. Without loss of the generality, we can assume that $\alpha$ is a positive real number, and $\beta=|\beta|e^{i\theta}$. We aim to create a target entangled state where the photon is conditionally reflected if the flux qubit is in the excited state $|\Phi_T(x, t\rightarrow \infty)\rangle = \overline{\phi}_R(x) e^{i\varTheta(x)/2}C_\text{R}^\dag(x) |\varnothing,g\rangle - \overline{\phi}_L(x)e^{-i\varTheta(x)/2} e^{i\varphi}C_\text{L}^\dag(x) |\varnothing,e\rangle$, where $\varTheta(x)$ is a spatial phase factor due to the opposite propagation directions of the two parts of the propagating wave functions. We note that the phase $\varTheta$ does not appear when taking the absolute of the overlap  between the wave packet components defined as $|1_R\rangle =: \bar{\bar{\phi}}_R(x)C_\text{R}^\dag(x)|\varnothing\rangle$ and $|1_L\rangle =: \bar{\bar{\phi}}_L(x) C_\text{L}^\dag(x)|\varnothing\rangle$, which is a measure of  coherence, while $\varphi$ is a trivial small phase offset. Here we choose $\int |\bar{\bar{\phi}}_R(x)|^2dx=1$ and $\int |\bar{\bar{\phi}}_L(x)|^2dx=1$. $|1_R\rangle$ means a single photon in the right-moving mode over the whole waveguide, while $|1_L\rangle$ is for a left-moving single photon. Ideally, we have $\int |\bar{\phi}_R(x)|^2 dx = |\alpha|^2$ and $\int |\bar{\phi}_L(x)|^2 dx = |\beta|^2$. So, we have $\bar{\phi}_R(x)=\alpha\bar{\bar{\phi}}_R(x)$ and $\bar{\phi}_L(x)=\beta \bar{\bar{\phi}}_L(x)$. We define the overlap fidelity with a target state as 
\begin{equation} \label{eq:F}
 \begin{split}
  F(\varphi) = & \int dx Tr[\rho |\Phi_T(x)\rangle \langle \Phi_T(x)|] \\
             = & |\alpha|^2 \int dt Tr[\rho(t) C_\text{R}^\dag|\varnothing,g\rangle \langle \varnothing,g| C_\text{R}] \\
             & + |\beta|^2 \int dt Tr[\rho(t) C_\text{L}^\dag|\varnothing,e\rangle \langle \varnothing,e| C_\text{L}] \\
               & + \alpha \beta^* \int dt e^{i\varTheta(t) +i\varphi} Tr[\rho(t) C_\text{R}^\dag|\varnothing,g\rangle \langle \varnothing,e| C_\text{L}] \\
               & + \alpha^* \beta \int dt e^{-i\varTheta(t) -i\varphi}Tr[\rho(t) C_\text{L}^\dag|\varnothing,e\rangle \langle \varnothing,g| C_\text{R}] \;.
 \end{split}
\end{equation}

The coherence at position $x$ is given by $\mathscr{C}(x)=  -e^{i\varTheta(x)+i\varphi}Tr[\rho C_\text{R}^\dag|\varnothing,g\rangle \langle \varnothing,e| C_\text{L}]$. This definition removes the fast oscillating phase between the two opposite propagating paths. The total coherence for the whole hybrid entangled state is evaluated as $\mathscr{C}_\text{sys} = \int dx \mathscr{C}(x)$.

To prove the entanglement between the output photon and the flux qubit, we also calculate the concurrence, $C$, in the basis of $\{|1_R,g\rangle, |1_R,e\rangle, |1_L,g\rangle, |1_L,e\rangle \}$. We first calculate the reduced density matrix of the moving photon and the static flux qubit,
$\rho_\text{ph-flux}(t) = \sum_{A,B,l,k} Tr\{ \rho(t) C_A^\dag |l\rangle \langle k| C_B\} |1_A,l\rangle \langle k, 1_B| $ with $A,B\in \{R,L\}$ and $l,k \in \{e,g\}$, by partial tracing over the SiV. Once we know this density matrix in this basis, we can calculate the concurrence involving the whole moving wave packets by the method developed by Wootters \cite{Concurrence} as following steps: (i) Find the time-dependent spin-flipped density matrix $\tilde{\rho}_\text{ph-flux}(t) = (M \otimes M) \rho^*_\text{ph-flux}(t) (M \otimes M) $, where $\rho^*_\text{ph-flux}(t)$ is the complex conjugate of $\rho_\text{ph-flux}(t)$, and $M = \begin{pmatrix} 0 & -i \\ i & 0 \end{pmatrix}$ takes the standard time reversal operation on a spin-$\frac{1}{2}$ particle. (ii)Calculate the time-dependent Hermitian matrix $R(t) = \sqrt{\sqrt{\rho_\text{ph-flux}(t)}\tilde{\rho}_\text{ph-flux}(t) \sqrt{\rho_\text{ph-flux}(t)}}$. (iii) Solve the eigenvalues $\lambda_j$ ($j\in \{1,2,3,4\}$), in decreasing order, of Hermitian matrix $R(t)$ at $t$. (iv) Calculate the concurrence density $C(x) = max\{0, \lambda_1(x) - \lambda_2(x)- \lambda_3(x)- \lambda_4(x) \}$ at $x$. Here we use the fact that $x=v_g t$ and set $v_g=1$ again. (v) The total concurrence for the whole wave packets is evaluated as the integral of concurrence density, i.e. $C= \int dx C(x)$.  
We simply write the above calculation for the concurrence as $C = \int C[\rho_\text{ph-flux}(t)] dt$.

The method developed by Shanhui Fan et al. can easily derive the analytic form for the steady-state solution of the system without any dephasing effects arising from the relaxation and pure dephasing. In contrast, the master equation model can provide a full numerical solution of the system with taking into account the dephasing. The master equation method can completely recover the steady-state transmission and reflection derived from Fan's method by using a long enough input pulse. These two method are equivalent in solving the steady-state solution for a single photon input if the dephasing is negligible. The analytic formula can be useful for finding a working window.

\subsection{Available coupling strength}
\begin{figure}
 \centering
 \includegraphics[width=0.99\linewidth]{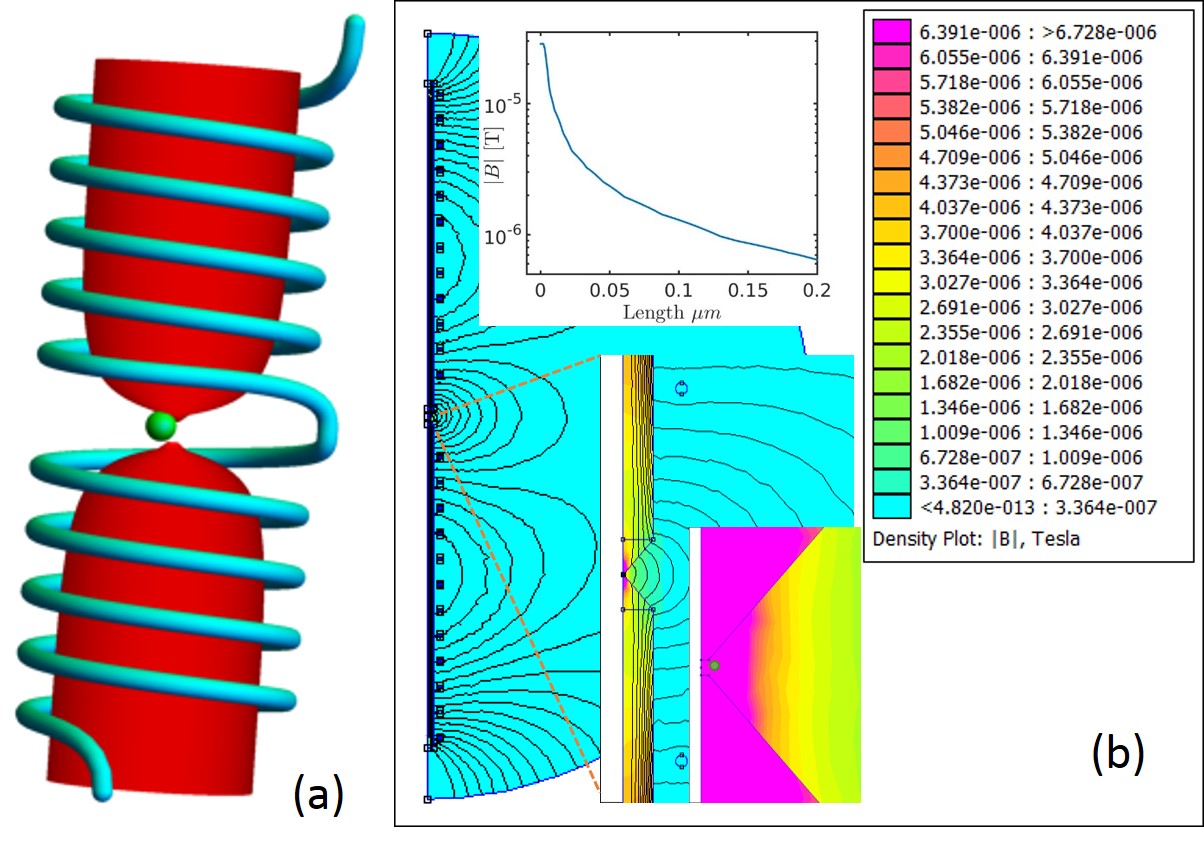} \\
 \caption{(Color online) Estimation of the magnetic field at the location of SiV generated by the flux qubit. (a) Schematic of the coiled part of flux qubit for enhancing the magnetic field. The flux qubit has a $24$-turn coiled edge (Cyan tube). A mu-metal bowtie flux concentrator (red tapered cylinder) with a relative permeability of $\mu_r=10^6$ is inserted inside the coil to greatly enhance the local magnetic field generated by the flux qubit. The SiV (green ball) locates nearby the tip of concentrator. (b) The distribution of magnetic field around the structure in (a) is numerically solved in two-dimensional space with the software FEMM4.2. The density plot inset shows the zoomed-in distribution nearby the concentrator tips, while the line plot inset shows the distribution along the middle line of two tips. The small green ball in the inset indicates the position of SiV.} \label{fig:Bfield}
\end{figure}

Before we start to study the motion of the quantum system, we numerically estimate the magnetic field generated by the flux qubit at the location of the SiV with the software FEMM 4.2 version. Figure \ref{fig:Bfield}(a) depicts the structure for enhancing the magnetic field with a mu-matel bowtie-shaped flux concentrator. One edge of the flux qubit is $24$-turn coil with a bowtie flux concentrator inside. We assume that the wire of flux quit has a radius of $50~\text{\nano\meter}$ and the persistent current leading along the flux qubit is $500~\text{\nano\ampere}$. The gap between two turns is $1~\text{\micro\meter}$. 
The flux concentrator is made from mu-metal with a relative permeability of $\mu=10^6$ \cite{BHose}. Such bowtie structure has been demonstrated to be capable of focusing magnetic field efficiently \cite{BConcentrator}. The larger end of bowtie is cylinder with radius of $260~\text{\nano\meter}$. The tips of bowtie end have a radius of $5~\text{\nano\meter}$. The gap between two tips is $5~\text{\nano\meter}$.  $5~\text{\nano\meter}$ away from the center of bowtie the magnetic field can be $\sim 21~\text{\micro\tesla}$ yielding $\eta/2\pi\sim 300~\text{\kilo\hertz}$. The coupling  is strong enough for our single-photon router here. The closet turn of the coil is a few $\micro\meter$ away from the SiV. Such design has an advantage over the state-of-the-art design using a $15~\text{\nano\meter}$ superconducting nanowire \cite{LargeBCoupling}. It can isolate the SiV from the superconducting wire to avoid the loss of superconductivity of the flux qubit. We assume that the nanowaveguide is a short plasmonic transmission line \cite{LargePurcellPlasmon} so that the SiV and the nanowire can be inserted into the free space of the bowtie tip.

\section{Results}
\subsection{Working window}
Now we determine the dependence of the transmission and reflection of a single-photon pulse on the quantum state of the flux qubit.
\begin{figure}
 \centering
 \includegraphics[width=0.6\linewidth]{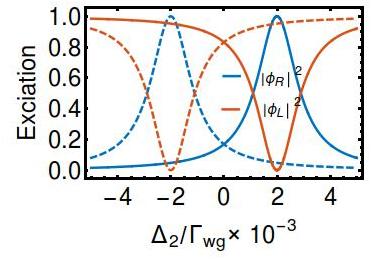} \\
 \caption{(Color online) Steady-state transmission (blue lines) and reflection (red lines) for the excited state (dashed lines) and ground state (solid lines) of the flux qubit. We set $\Omega_c/\Gamma_\text{wg}=0.03, \eta/\Gamma_\text{wg}=10^{-3},\Delta_1=0, \gamma_f=0, \Gamma^*=0, \xi_c=1, \xi_2=0.5$.} \label{fig:workingwindow}
\end{figure}
To show the main idea we first neglect the decay and decoherence of the system by setting $\gamma_f=0, \Gamma^*=0$.
Choosing appropriate parameters we find a working EIT window for a small $\eta=10^{-3}\Gamma_\text{wg}$, see Fig. \ref{fig:workingwindow}.  From this we observe that the quantum state of the flux qubit can control well the propagation of a single-photon pulse  around $|\Delta_2|=2|\eta|$, with a bandwidth of $\Delta \omega = 2\Omega_c^2/\Gamma_\text{wg}$, if the EIT window is narrow, e.g. $\Omega_c = 0.03\Gamma_\text{wg}$.

\subsection{Single Photon Wavepacket}
\begin{figure}
 \centering
 \includegraphics[width=0.98\linewidth]{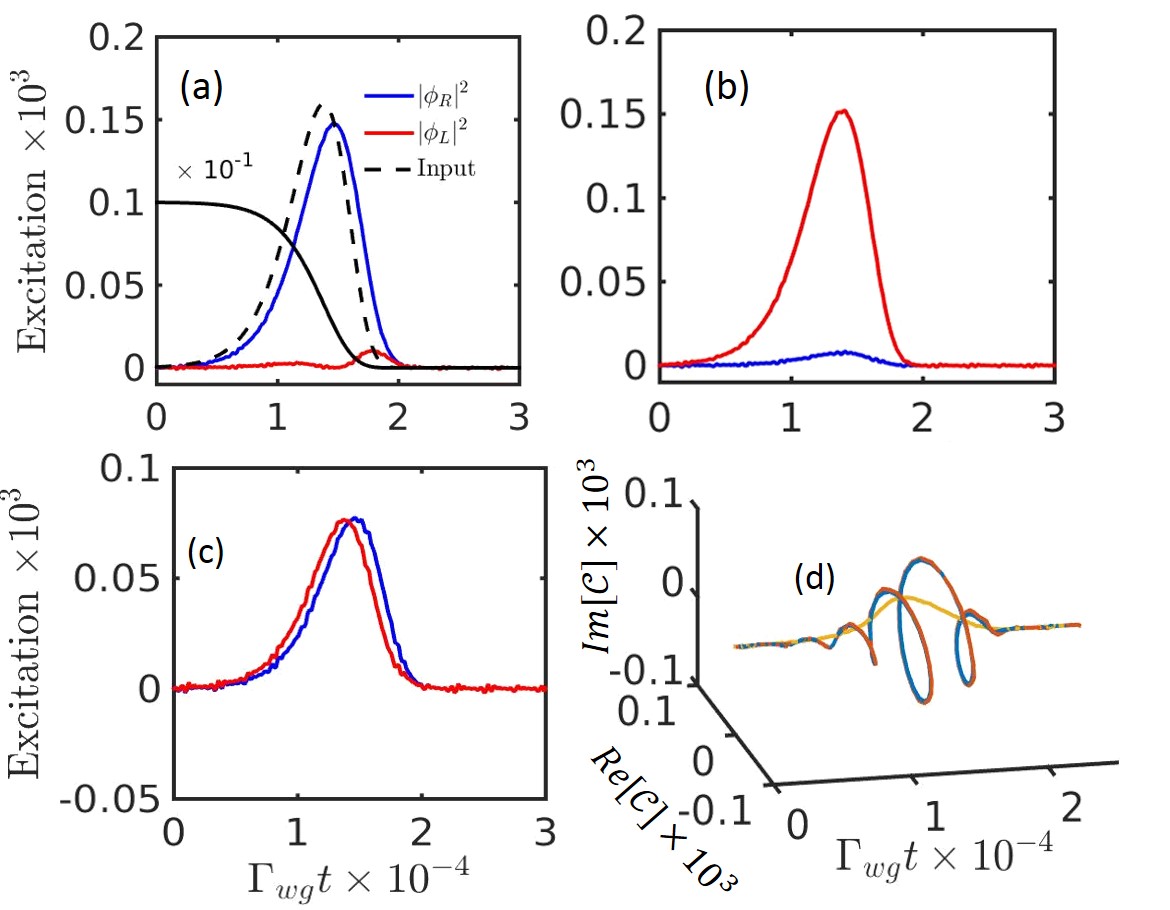} \\
 \caption{(Color online) (a-c) Time-evolution of excitations of right-(blue lines) and left (red lines)-moving photons with the initial state of the flux qubit (a) in $|g\rangle$, (b) in $|e\rangle$ and (c) in $(|g\rangle + |e\rangle)/\sqrt{2}$. (d) The coherence for the flux qubit prepared in $(|g\rangle + |e\rangle)/\sqrt{2}$. The blue solid line shows $Tr[\rho C_\text{R}^\dag|\varnothing,g\rangle \langle \varnothing,e| C_\text{L}]$ indicating the coherence density modulated by the fast spatial (time) oscillating phase $\varTheta(x)$. It is fitted by $-|\overline{\phi}_R(x)| |\overline{\phi}_L(x)| e^{i(\eta_1 + \eta_3)x +i\varphi+i\theta}$ (the purple line). The yellow line is for the real coherence evaluated by $\mathscr{C}(x)$. In (a), the dashed black line shows the input pulse and the solid black line is for the time-dependent excitation in the source cavity. Other parameters are $\Omega_c/\Gamma_\text{wg}=0.03, \eta/\Gamma_\text{wg}=10^{-3}, \Delta_1=0,\Delta_2=2\eta, \gamma_f=0, \Gamma^*=0, \xi_c=1, \xi_2=0.5$ and $\tau_p\Gamma_\text{wg}=10^4$. $\Gamma_\text{wg}/2\pi \approx 300~\text{\mega\hertz}$ for SiV.} \label{fig:Prop}
\end{figure}

Below we investigate the right- and left-moving wave packets for an input single-photon pulse, see Fig. \ref{fig:Prop}. The excitation of the ``source'' cavity first decays slowly and then quickly decays to zero. As a result, the input pulse totally includes only one excitation, and has a finite duration of $\tau_p \Gamma_\text{wg}=10^{4}$, which is within the bandwidth of the right hand working window in Fig. \ref{fig:workingwindow}. When the flux qubit is prepared in $|g\rangle$ corresponding to $\delta=0$, see Fig. \ref{fig:Prop}(a), the input pulse mostly passes through the SiV, $e_R\simeq 0.95$, and the reflection is very small, $e_L\simeq 0.05$, yielding a contrast of $\mathfrak{C}=0.91$. For the state $|e\rangle$ yielding $\delta=-4\eta$, see Fig. Fig. \ref{fig:Prop}(b), the single-photon pulse is mostly reflected backward, while the transmission is vanishing small. In this case, we have $e_R\simeq 0.05$ and $e_L\simeq 0.95$ corresponding to $\mathfrak{C}=0.91$. More interestingly, the input single photon is routed into a path-entangled state of the right- and left-moving wave packets with even excitations $e_R\simeq e_L\simeq0.5$, see Fig. Fig. \ref{fig:Prop}(c), if the flux qubit is in the superposition state $(|g\rangle + |e\rangle)/\sqrt{2}$. In the absence of the decoherence of the flux qubit, we create the entangled state of $|1_R,g\rangle$ and $|1_L,e\rangle$ in the bipartite system of the flying photon and the static flux qubit. The fidelity is about $F(0.07\pi)=0.943$ and the concurrence is $C=0.92$. 
The coherence between $|1_R,g\rangle$ and $|1_L,e\rangle$ is shown Fig. \ref{fig:Prop} (d). The evaluation of $\mathscr{C}_1(x)=  Tr[\rho C_\text{R}^\dag|\varnothing,g\rangle \langle \varnothing,e| C_\text{L}]$. It indicates the coherence density between $C_\text{R}^\dag |\varnothing,g\rangle$ and $C_\text{L} |\varnothing,e\rangle$ modulated by a fast oscillating phase $\varTheta(x)$ caused by the spatial phases of the opposite propagating wave functions. As a result, it is a spiral line (see the blue line). Our numerical simulation shows that $\varTheta(x) = (\eta_1+\eta_3)x$ and $\varphi=0.07\pi$ in $|\Phi_T\rangle$. Based on this finding, we fit $\mathscr{C}_1(x)$ with $-|\overline{\phi}_R(x)| |\overline{\phi}_L(x)| e^{i(\eta_1 + \eta_3)x +i\varphi+i\theta}$ (see the purple line). The real coherence $\mathscr{C}(x)$, removing the fast spatial oscillating phase, is positive real and resembles $|\overline{\phi}_R(x)\overline{\phi}_L(x)|$ (see the yellow line). The total coherence is $\mathscr{C}_\text{sys} \approx 0.466$. Obviously, the transmitted and reflected pulses resemble the input pulse in all cases.
\begin{figure}
 \centering
 \includegraphics[width=0.6\linewidth]{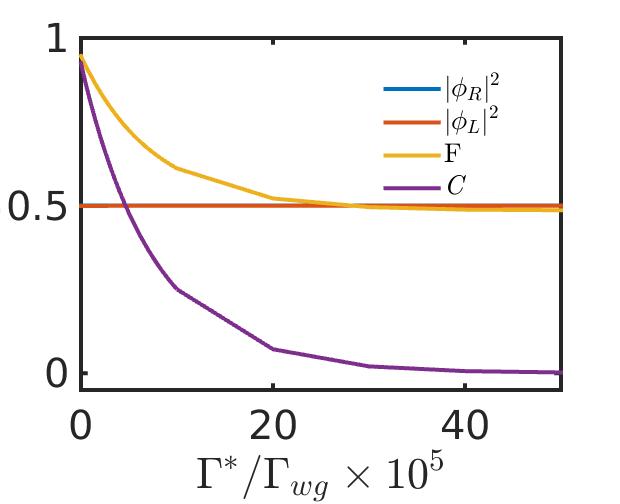} \\
 \caption{(Color online) Excitations ($\left| \phi_R \right|^2$, $\left|\phi_L\right|^2$), fidelity, $F$, and concurrence, $C$, as a function of the pure dephasing rate $\Gamma^*$. Other parameters are $\Omega_c/\Gamma_\text{wg}=0.03, \eta/\Gamma_\text{wg}=10^{-3}, \Delta_1=0,\Delta_2=2\eta, \gamma_f=0, \xi_c=1, \xi_2=0.5$ and $\tau_p\Gamma_\text{wg}=10^4$. The blue line ($|\phi_R|^2$) overlaps with that for $|\phi_L|^2$. $\Gamma_\text{wg}/2\pi \approx 300~\text{\mega\hertz}$ for SiV.} \label{fig:FCE}
\end{figure}

The entangled state $|\Phi_T(x)\rangle$ is the most interesting result of this work. It is fragile to the decay and  decoherence. The decay rate of the flux qubit can be reduced to be very small by decoupling it from the TL by modulating the terminating SQUID. However, the qubit's pure dephasing, $\Gamma^*$, will not reduce even when decoupled from the TL. 
Now we study the effects of the flux qubit pure dephasing and observe (see Fig. \ref{fig:FCE}), as  pure dephasing increases, the fidelity to the ideal target entangled state exponentially decays to that of a completely mixed state, $F=0.5$. Similarly, the concurrence also exponentially decays to zero and the entanglement disappears. In contrast, the right and left moving photon excitation probabilities are equal $0.5$ and independent of dephasing. Thus, for large dephasing  the system classically randomly routes the single photon into a mixture of right- and left-moving modes.

\subsection{Heralded quantum state transfer}
Next we show how to transfer quantum state between the ``flying'' photonic qubit and the ``static'' flux qubit with the help of quantum measurement. As discussed above, for a right-moving single photon and the flux qubit prepared initially in the state of $|\Psi_\text{f,in}\rangle=(\alpha |g\rangle + \beta  |e\rangle)$, our hybrid quantum system can generate an entangled state $|\Phi_T(x)\rangle = \alpha |1_R,g\rangle - \beta |1_L,e\rangle$. Here we neglect the small phase $\varphi$ and the spatial dependent phase $\varTheta(x)$, which only indicates the opposite propagation directions. We rewrite the state $|\Phi_T(x)\rangle$ as $|\Phi_T(x)\rangle= (\alpha |1_R\rangle - \beta |1_L\rangle) |\Psi_+\rangle - (\alpha |1_R\rangle + \beta |1_L\rangle)|\Psi_-\rangle$ in the Bell state basis of the flux qubit, where $|\Psi_\pm\rangle = (|e\rangle \pm |g\rangle)/\sqrt{2}$. Clearly, heralded quantum state transfer can be conducted by measuring the state of the ``static'' flux qubit in the Bell basis. A measurement yielding $|\Psi_-\rangle$ projects the ``flying'' photonic qubit to $(\alpha |1_R\rangle + \beta |1_L\rangle)$. Here we neglect the travail global phase of $\pi$. The success probability can be $50\%$. If we obtain $|\Psi_+\rangle$ during the measurement, we need induce a $\pi$ phase shift in either path of the photon. Thus, we can do a measurement-based heralded quantum state transfer from the ``static'' flux qubit to the ``flying'' photonic qubit.

\section{Experimental Implementation} 
We now provide an  estimate for the performance of our device for entangling the traveling single photon and the flux qubit, prepared in the superposition state of $(|g\rangle + |e\rangle)/\sqrt{2}$, based on feasible experimental parameters. The decay rate $\gamma_2$ of a single SiV is about $2\pi \times 30 ~\mega\hertz$ corresponding to a lifetime of $\sim 3~\nano\second$ \cite{FedorSiVEIT}. For simplicity, we use $\gamma_1=\gamma_2$ and $\Gamma=0$. The applied single-photon pulse has a duration of $\tau_p=30 ~\micro\second$.
The challenging requirements in the realization of our system are three-fold: (i) large single-spin Zeeman shift; (ii) the long decoherence time of the flux qubit; (iii) a strong coupling between the nanowaveguide and the atom yielding a Purcell factor larger than $10$. Using a flux qubit with a persistent current of $500~\text{\nano\ampere}$ and a flux concentrator we can achieve a coupling strength of $\eta/2\pi=300 ~\kilo\hertz$. In a recent experiment \cite{FluxQubit}, the flux qubit exhibited a longitudinal lifetime of $T_1\sim 44 ~\micro\second$ and a decoherence time of $T_2 \approx 80 ~\micro\second$ giving a pure dephasing rate $T_2^* \approx 880 ~\micro\second$ \cite{SQreview1}. Thus we have $\Gamma^* \approx 2\pi\times 181 ~\hertz$. Since the transmission line is terminated by a SQUID we assume that through appropriate tuning we can eliminate all longitudinal relaxation of the flux qubit, giving  $\gamma_f=0$ during the single photon pulse. Recent progress in nanowaveguide QED has achieved the strong coupling regime with a Purcell factor larger than $20$ through a variety of methods: using a nanofiber \cite{LargePurcellNanofiber}, or a dielectric slot nanowaveguide \cite{LargePurcellSlotwg}, or a photonic crystal waveguide \cite{LargePurcellPCwg1,LargePurcellPCwg2,LargePurcellPCwg3,LargePurcellPCwg4,LargePurcellPCwg5} or a plasmonic nanowire \cite{LargePurcellPlasmon}. We use a Purcell factor of $\Gamma_\text{wg} / \gamma_2 = 10$ to achieve the strong coupling regime in our estimation. Using these realistic numbers for parameters in the master equation model, we achieve a high fidelity of $F=0.87$ and entanglement with a large concurrence of $C=0.83$. Throughout our numerical investigation, we require $\Omega_c=0.03\Gamma_\text{wg}=2\pi\times 9~\text{\mega\hertz}$. Considering the large dipole moment of SiV, this classical Rabi frequency can be reached with a weak laser field.

\section{Conclusion}
In summary, we can conditionally control the routing of a single photon wavepacket by the quantum state of a flux qubit via  quantum magnetic tuning of the position of an EIT window in a single SiV color center interacting strongly with the single photon. Our scheme can create a quantum state of a flying single photon dependent on the quantum state of a flux quibt. 
The proposed device can act as a hybrid quantum interface and creates entanglement between the mw and optical regimes. 
The atom in the implementation of our device can be, but is not limited to
the SiV defect. It can be NV centers in nanodiamond, rare
earth ions in nanoscrystal or alkali atoms.

\section*{Funding.}
This research was supported in part by the ARC Centre
of Excellence in Engineered Quantum Systems (EQuS),
Project No. CE110001013. FJ also would like to thank the SIQS EC Project No. 600645  for support.



\end{document}